# Evidence for a liquid precursor to biomineral formation.

*Cayla A. Stifler[1], Christopher E. Killian[2], Pupa U. P. A. Gilbert[1,3,4†*].*

Cayla A. Stifler[1], Christopher E. Killian[2], Pupa U. P. A. Gilbert[1,3,4†*].

1 Department of Physics, University of Wisconsin, Madison, WI 53706, USA.

2 Department of Molecular and Cell Biology, University of California, Berkeley, CA 94720, USA.

3 Departments of Chemistry, Geoscience, Materials Science, University of Wisconsin, Madison, WI 53706, USA.

4 Lawrence Berkeley National Laboratory, Chemical Sciences Division, Energy Geoscience Division, Berkeley, CA 94720, USA.

† previously publishing as Gelsomina De Stasio.

STATEMENT OF URGENCY AND BRIEF SUMMARY

The observation of a liquid precursor in sea urchin spines significantly expands the number of mineral phases and phase transitions involved in biomineral formation. With this observation there are 5 phases and 4 phase transitions, including: liquid-liquid phase separation, precipitation, dehydration, crystallization. At present, most people have no hope that isolating liquid precursors is even possible, thus they will not even try, given the past 20 years of failures. The rapid publication of this communication will effectively and quickly jump-start a new field of investigation.

ABSTRACT

The crystals in animal biominerals such as sea urchin spines, mollusk shells, and coral skeletons, form by attachment of amorphous particles that subsequently crystallize. Do these solid amorphous precursor particles have liquid precursors? Polymer-induced liquid precursors (PILP), or prenucleation clusters coalescing into a liquid precursor to calcium carbonate crystallization have been observed extensively in synthetic systems. Molecular dynamics simulations also predict liquid-liquid phase separation. However, evidence for liquid precursors in natural biominerals remains elusive. Here we present Scanning or PhotoEmission Electron Microscopy (SEM, PEEM) evidence consistent with a dense liquid-like precursor in regenerating sea urchin spines. The observed precursor originates in tissue and ultimately transforms into a single crystal of calcite ($CaCO_3$) with complex stereom morphology.

MAIN TEXT

Animal biomineralization played a central role during the evolutionary history of animals [1-2] by providing a selective advantage over soft-bodied species. Biominerals have advantageous material properties [3], different from geologic minerals [4]. The study of how animals construct mineralized structures is a compelling materials, cell, and evolutionary biology problem. Many previous studies of biomineralization in animals have found that amorphous precursors [5-6] and crystal formation by attachment of amorphous precursor particles [7] are vital components in forming calcium carbonate biominerals.

Liquid precursors to biomineral formation, long suspected to exist [8], have been elusive. Such liquid precursors are observed in synthetic systems [9-13], but not naturally occurring biominerals despite extensive searches [14]. Liquid precursors remain an unobserved assumption in 2021 [12].

Here we present imaging and spectromicroscopic evidence of a liquid-like material in regenerating sea urchin spines, revealed by Scanning and PhotoEmission Electron Microscopies (SEM, PEEM). At the time of observation with both methods, the liquid-like material is solid. In SEM its chemical composition is similar but not identical to crystalline calcite ($CaCO_3$), in PEEM it is more amorphous than any other part of the forming spine. The liquid-like material pools seen in SEM have a smooth surface similar to a liquid. In particular, the liquid-like material stands out where typically empty holes are present in the sea urchin spine stereom, but it can also coat the surface of the stereom. The stereom is an intricate calcite structure, with intersecting 10-$\mu$m-wide cylindrical struts and 10-$\mu$m-diameter holes common to all echinoderm skeletal parts [15-16]. The stereom diffracts like a single crystal of calcite [17], but, unlike abiotic calcite crystals, it does not have a thermodynamically favorable euhedral morphology, with flat faces and sharp edges. Instead, the stereom has smooth, curved surfaces. Topologically the stereom is a gyroid, that is, an infinitely connected triply periodic structure that minimizes surface area and does not contain straight lines [18]. The present study focuses on examining the nature of the calcium carbonate precursors and observing liquid phases, if they exist, in regenerating sea urchin spines.

When calcitic sea urchin spines are broken in nature or cut in the laboratory, they regenerate [19]. The regenerating part is well-distinct, as it starts growing from the center of the cylindrical spine and thus has a smaller diameter than the original spine. The color is also distinct: deep purple in the original spine of the sea urchin species studied here and pale brown in the regenerating spine. Fig. 1a shows spines cut using scissors and photographed approximately 5 days later in a living California purple sea urchin, *Strongylocentrotus purpuratus*. Seven to 14 days after cutting them, the spines re-grow to the original size near the cutting position and taper in diameter towards the regenerating tip. The cut spine is an ideal system to observe calcite formation mechanisms. Adult sea urchin spines, along with sea urchin larval spicules and mollusk larval shells, were among the first biominerals in which amorphous calcium carbonate (ACC) precursors were observed [5-6, 20].

A simple sample preparation revealed the liquid-like precursor in Fig. 1c-f. The preparation is entirely done on ice and includes rinsing in Tris base at pH 9, bleaching, rinsing in Tris base, and dehydrating in ethanol. The crucial difference between this and previous protocols is that the spines were kept vertical, with the regenerating tip pointing up during all preparation steps. Bleach oxidizes and removes the tissue around the spine stereom. With the spines upright and their regenerating tips up, precursor phases should fall and accumulate on top of the stereom and are less likely to wash away during rinsing and washing. The unusually dense mineral observed in Fig. 1c-f covering the stereom, therefore, was not washed away but remained on the stereom and filled the stereom holes, behaving like a dense liquid. At the time of observation with SEM in vacuum, the material was likely no longer liquid. Still, it retained the morphology of liquid materials, filling all holes in a region that should otherwise show plentiful stereom holes. Fig. S1 shows fractured spines, which, at their centers, only have stereom [19]. Thus, the location of the liquid-like material near the center of a spine's stereom provides an ideal substrate to distinguish solid from liquid materials. The SEM images in Fig. 1 were acquired in back-scattered electron (BSE) mode; thus, their brightness is proportional to the atomic number of the elements observed. The liquid-like material is as bright as the calcite stereom; therefore, it must have similar composition and density. Energy-dispersive x-ray (EDX) fluorescence emission data in Fig. S2 show that the two spectra are similar but not identical, confirming that the liquid-like pool has the same relative amounts of Ca and O as calcite in the mature spine, but substantially more C. Greater C concentration in liquid-like pools than in mature stereom is consistent with organic molecules stabilizing the liquid, as they do with synthetic PILP [8].

In Fig. 3, PEEM component maps show a regenerating spine with many pixels in the stereom still amorphous, as well as in-tissue particles. These particles have average size ~200 nm, and are located in the tissue, ~5-10$\mu$m away from the stereom. This distance is not surprising, as calcifying cells are known to transport calcium through filopodia that are 5-10$\mu$m long [21]. Three areas on the same spine were analyzed in total, as shown in Fig. 3, containing a total of 13 in-tissue particles. To quantify how disordered the in-tissue phases were, we used the method recently developed by Kahil et al. for a similar biomineral-forming system: intracellular Ca-rich particles in the cells that deposit calcite spicules in sea urchin embryos [22]. We found three phases, calcite, ACC, and ACC-$H_2O$ (Fig. S3), in the in-tissue particles of three areas of the regenerating spine (see Fig. S4 for their precise location), as were observed in sea urchin embryo cells by Kahil et al. [22]. In-tissue particles are much more amorphous (i.e., contain more red and yellow pixels) compared to the stereom, which contains mostly blue or cyan pixels (Fig. 3). Peak-ratio analysis confirms that in-tissue particles are more disordered (Fig. S3). This observation is consistent with such particles being more disordered than the most disordered solid phase, ACC-$H_2O$, and thus were not completely solid during PEEM analysis.

We hypothesize that liquid droplets formed within spine calcifying cells, explaining why they can be more disordered than ACC-$H_2O$ in the in-tissue particles at the time of PEEM analysis. When the tissue was bleached away for SEM analysis, the droplets from many cells coalesced and formed the larger pool observed in Fig. 1c-f. Such intracellular Ca-rich, membrane-bound particles or droplets were directly observed by Kahil et al. in single cells [22].

We prepared a total of 28 regenerating spines from 9 different sea urchins, all *Strongylocentrotus purpuratus*. Pools of liquid-like material were observed consistently in 20 spines, with the one in Fig. 1c-f being the largest pool and many smaller pools observed to occlude single holes in the stereom. Also, at the time of observation in the SEM, these were probably mostly solid, but they retained the morphology consistent with a liquid surface in a pseudomorphic transformation. Fig. 2 shows two liquid-like pools, demonstrating that they are smooth at the nanoscale and that they are as bright as the surrounding calcite stereom in both secondary electrons (SE-) and BSE-SEM mode.

After vertical-spine-bleaching preparations, in addition to the liquid-like pools, we also observe calcite rhombohedra, which are never observed in sea urchin spines bleached horizontally. One possible interpretation is that these rhombohedra formed at the expense of either the liquid precursor or the ACC solid precursor particles known to exist in regenerating sea urchin spines [19]. The rhombohedra are presented in Fig. S5 and Fig. 2b3,b4. The rhombohedra are as bright as the stereom in BSE-SEM (Fig. 2b4, Fig. S6a4), whereas cubic crystals of salt are bright in SE and dark in BSE (Fig. S6d4). We thus conclude that the bright euhedral crystals in Fig. S5 are calcite rhombohedra. The possibility that the rhombohedra formed from the liquid precursor is supported by the morphology of the liquid-like material observed immediately adjacent to some rhombohedra, e.g., those in Fig. S5b4,c4 or Fig. S5b1,c1. These patches with dense-liquid morphology are unusual on the surface of the stereom, were never observed before, and only appear when using the present vertical-spine-bleaching preparation. Thus, the unusual rhombohedra are more likely to have formed from the unusual dense liquid-like material observed here than from dissolution and re-precipitation of the solid particles previously reported [19]. Similar thin layers of synthetic PILP and non-pseudomorphic transformations were observed by Gower et al. on glass slides or microporous molds shaped after a natural sea urchin spine [23-24].

Other clues to the existence of a liquid precursor are presented in Fig. 4. These include the observation that the thin liquid-like pools are ~100-nm thick (Fig. 4c4), at least at the time of observation in SEM, consistent with a liquid with high surface tension forming a layer in a hole, like soapy water in a soap bubble loop. Another observation, unique in this set of experiments, was a circular hole in a liquid-like pool, reminiscent of an inverted bubble, suggesting liquid-like behavior and cracking that is solid-like behavior (Fig. 4a3,a4). A possible interpretation is that the material was liquid first, then solidified and cracked during water removal and volume change. A gel or colloidal suspension of particles could also explain this solid-like behavior, but not thin soap bubble layers.

We did not observe the liquid-like material in all spines. In 8 out of 28 spines, only the usual, smooth, solid stereom was found. In one spine only, we observed solid particles filling the stereom holes (Fig. 4d2). A possible interpretation is that the liquid precursor solidified before reaching the stereom when the tissue was bleached, and the solid particles fell into the stereom holes.

We tentatively conclude that liquid droplets are precursors to the ACC precursor solid particles previously observed in sea urchin spine stereoms [19]. The whole set of phases and phase transitions involved in sea urchin spine biomineralization is presented in Table 1.

**Table 1.** Five phases and four phase transitions leading to the formation of crystalline calcite in sea urchin biominerals. Ions can attach to any of the phases and during all phase transitions.

| Phase | Chemical Formula |
|---|---|
| Aqueous solution | $H_2O$, $Ca^{2+}$, $CO_3^{2-}$, $HCO_3^-$, $Sr^{2+}$, $Mg^{2+}$, … |
| 1. Liquid-liquid phase separation ||
| Liquid precursor denser than water | $CaCO_3 \cdot (H_2O)_{n>1}$ |
| 2. Precipitation of ACC-$H_2O$ solid particles ||
| Solid precursor particles of hydrated amorphous calcium carbonate (ACC-$H_2O$) | $CaCO_3$-$(H_2O)_{n \approx 1}$ |
| 3. Dehydration ||
| Anhydrous amorphous calcium carbonate (ACC) | $CaCO_3$ |
| 4. Crystallization ||
| Calcite, aragonite, or vaterite | $CaCO_3$ |

The observation of liquid-like morphologies does not contradict previously observed ACC phases [5-6, 25], nor biomineral formation by particle attachment [7, 26-27]. Several previous studies showed calcein-labeled intracellular vesicles in embryonic cells depositing sea urchin spicules [21, 28-30]. Kahil et al. recently showed hundreds of intracellular vacuoles and vesicles that initially contain seawater, form ACC-$H_2O$, ACC, and calcite [22], the same phases previously observed with PEEM in sea urchin spine stereom [19] and in tissue here. We observe fewer in-tissue particles here compared to Kahil et al. [22] because the probing depth of PEEM is only 3 nm at the Ca L-edge [31], thus, the probability of finding Ca-rich particles in a 3nm-thick cell slice is much smaller than 6% (Fig. S7).

In all in-tissue particles, we found more amorphous phases than in the forming stereom, suggesting that the precursor phases are formed in the tissue and are then deposited on the growing stereom. In a few cases the spectra showed even greater disorder than the most disordered solid ACC-H$_2$O, which could be interpreted as not completely solidified (Fig. S3). Such disordered phase is consistent with a dense liquid phase, precursor to the solid ACC-H$_2$O and ACC phases, as listed in Table 1.

The liquid-like pools observed in SEM images cannot be solid particles originally suspended in a liquid because solid particles in sea urchin spines were observed to be 100-400 nm in size. Particles with such sizes would appear nanoparticulate at 50,000x, precisely the magnification at which solid particles are observed here in Fig. 4d4. Such particles were shown in sea urchin spines and many other biominerals by Gilbert et al. [26]. At this scale, all liquid-like pools observed here are smooth, not nanoparticulate.

The liquid precursor was denser than water. Otherwise, it would not have fallen onto the underlying stereom in the bleach-water solution. Even if the liquid droplets were formed in membrane-bound intracellular vesicles in the tissues and were not in contact with water, when the membranes dissolved in bleach, the liquid droplets did not mix with and disperse in water. Thus, a liquid-liquid phase separation (LLPS) must have occurred previously in intracellular vesicles. Alternatively, LLPS may occur in membrane-less intracellular assemblies, as described for proteins [32] and synthetic systems that replicate many of the biomineral features in vitro without vesicle membranes [11]. Since Kahil et al. observed membrane-bound vesicles containing Ca-rich particles in sea urchin embryonic cells [22], these are most likely membrane-bound in sea urchin spine cells as well.

LLPS leading to calcite formation was first observed in synthetic $CaCO_3$ systems by the Gower group[8, 11, 33], and then extensively reproduced by other groups in a variety of synthetic systems [9-10, 34-37]. LLPS was also predicted in molecular dynamics simulations [38]. The phase diagram of calcium carbonate solutions shows that at 15 °C, the temperature at which *S. purpuratus* sea urchins live, the spinodal line is at 2 mM, thus at this or greater concentrations, liquid-liquid separation can occur [39]. Seawater [Ca] is 10-11 mM, and in sea urchin intracellular vesicles [Ca] is much greater, 1-15 M [22], thus LLPS can occur.

The liquid precursor is likely common to other $CaCO_3$ biomineralizers. The same ACC-$H_2O$ and ACC solid precursors were observed in many fresh, forming biominerals from diverse animals. These biominerals include sea urchin spicules [25], teeth [40], and spines [19], in aragonite ($CaCO_3$) mollusk shell nacre [41] and coral skeletons [27, 42]. Thus, we expect other $CaCO_3$ biomineralizers to share a dense liquid precursor observed in sea urchin spines and the sequence of phases and phase transitions in Table 1. Further experiments on diverse biominerals will demonstrate or refute the robustness these preliminary results, and their generality to other systems. If confirmed, a liquid precursor would solve long-standing mysteries in nacre formation [43-44]. Twenty years after this idea was first proposed [8], and extensive searches [12], we present the best evidence to date for a liquid precursor to biomineral formation. At the time of analysis by SEM and PEEM, the liquid-like precursor had solidified, via a pseudomorphic transformation, that is, retaining the liquid-like morphology. Isolating the liquid precursor and measuring its viscosity and composition remains a challenge.

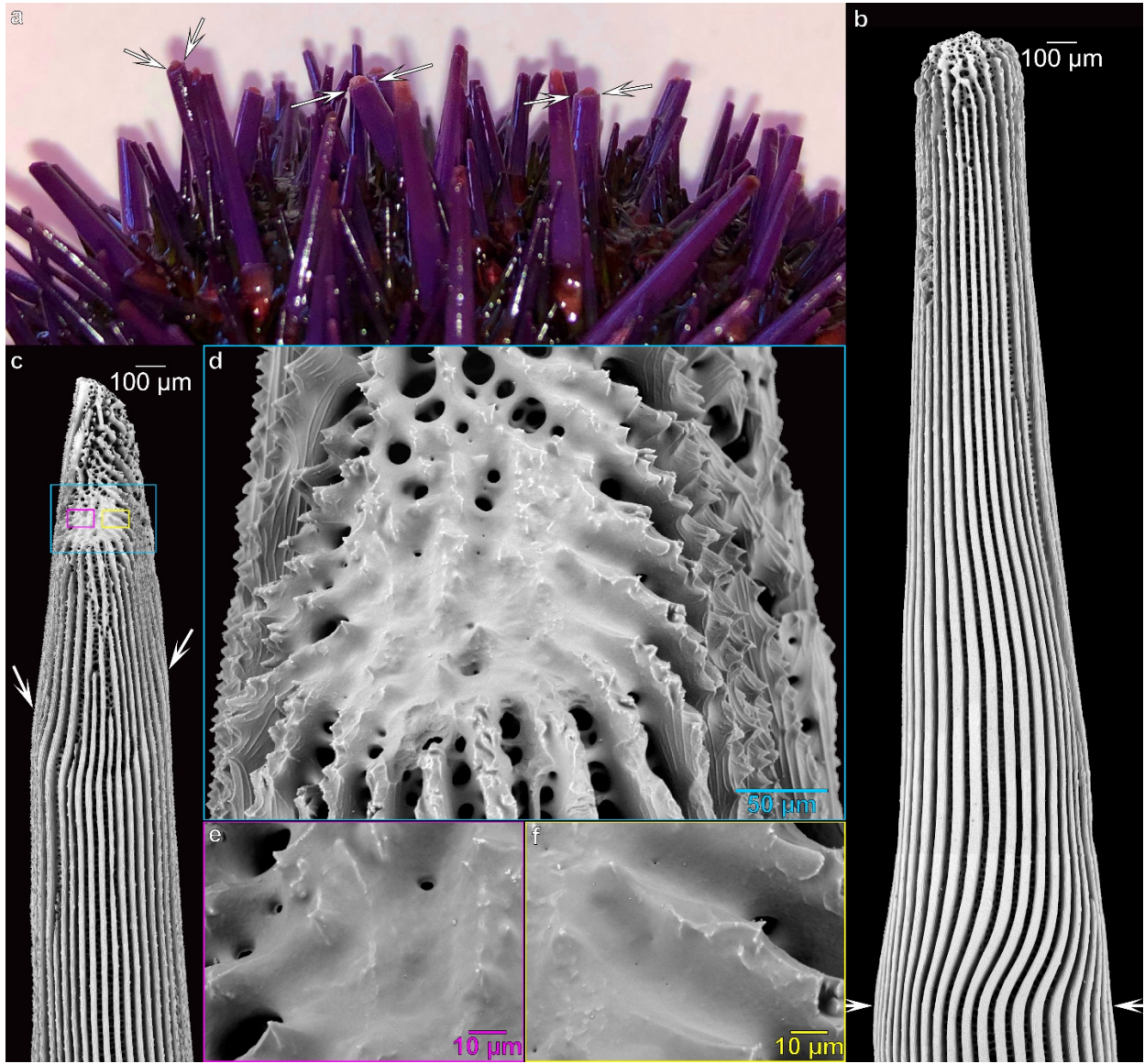

**Figure 1. Regenerating sea urchin spines show a liquid-like amorphous precursor**. a. Photograph of a living *Strongylocentrotus purpuratus* sea urchin, taken ~5 days after the spines were cut between arrowheads, obliquely between the arrowhead pair for the left spine and perpendicular to the spine axis between arrowheads pairs for the center and left spines. b, c. SEM images showing the original (below the arrows) and regenerating spine (above arrows) after ~10 days of re-growth. The spines were cut horizontally in b, and obliquely in c. In c, the original and regenerating spines are separated by a curved line between arrowheads, distinguishable because the spine diameter tapers down in the regenerating spine. Notice in c the liquid-like precursor filling stereom holes. d, e, f. Magnified images of the liquid-like precursor showing that it is smooth at the micron and sub-micron scale. Boxes in c indicate the regions magnified in d, e, f. All SEM images were acquired in back-scattered electron (BSE) mode.

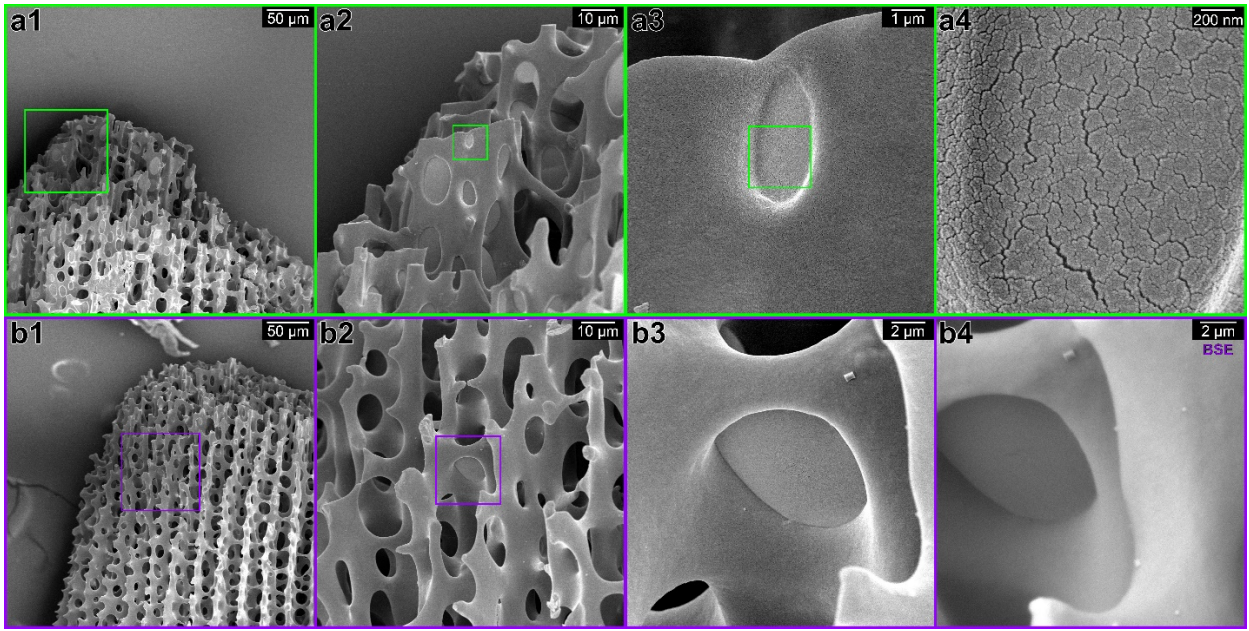

**Figure 2. Liquid-like pools in regenerating spines**. Increasing magnifications of the boxed regions are shown in subsequent panels, and correspondingly colored. a3 and b3 show two liquid-like pools in two spines from different animals. a4 shows that at high magnification the liquid-like pool is smooth and featureless, except for cracking in the Pt coating. a1-a2, b1-b2 show many more pools, as does Fig. S5. Notice that the pools in a3 and b3 are as bright as the rest of the stereom. b4 was acquired in back-scattered electron (BSE) mode, all other images in secondary electron (SE) mode. b3 and b4, showing the same pool in SE and BSE, confirm that the liquid-like pool has similar composition to the stereom.

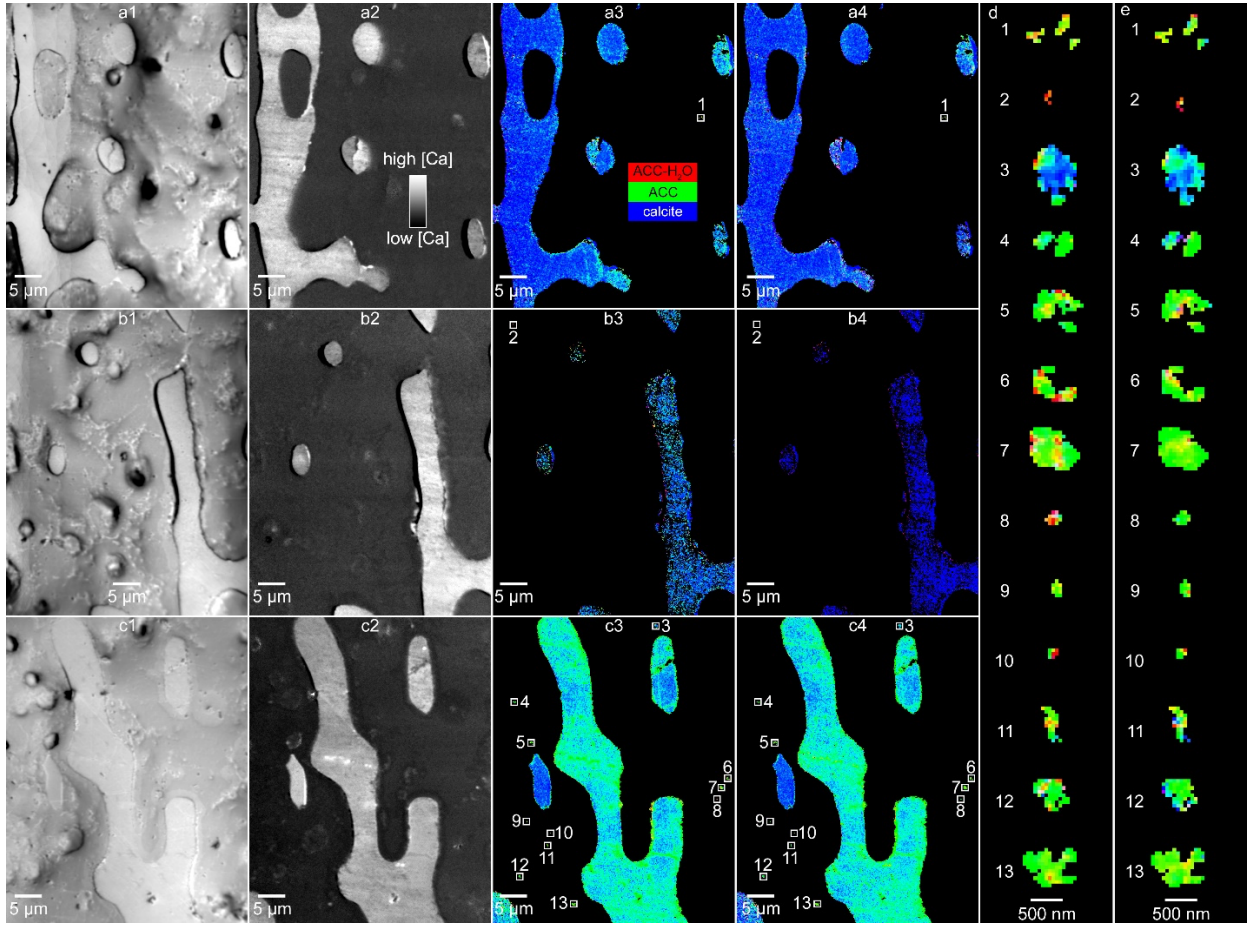

**Figure 3.** a1-c1 Average PEEM images, a2-c2 Ca distribution maps, a3-c3 PEEM-component maps, and a4-c4 repeat component maps of three different areas of a regenerating sea urchin spine. The numbered boxed regions in component maps a3-c3 and a4-c4 are magnified in d and e, respectively. In total, we found 13 different in-tissue particles in the three areas.

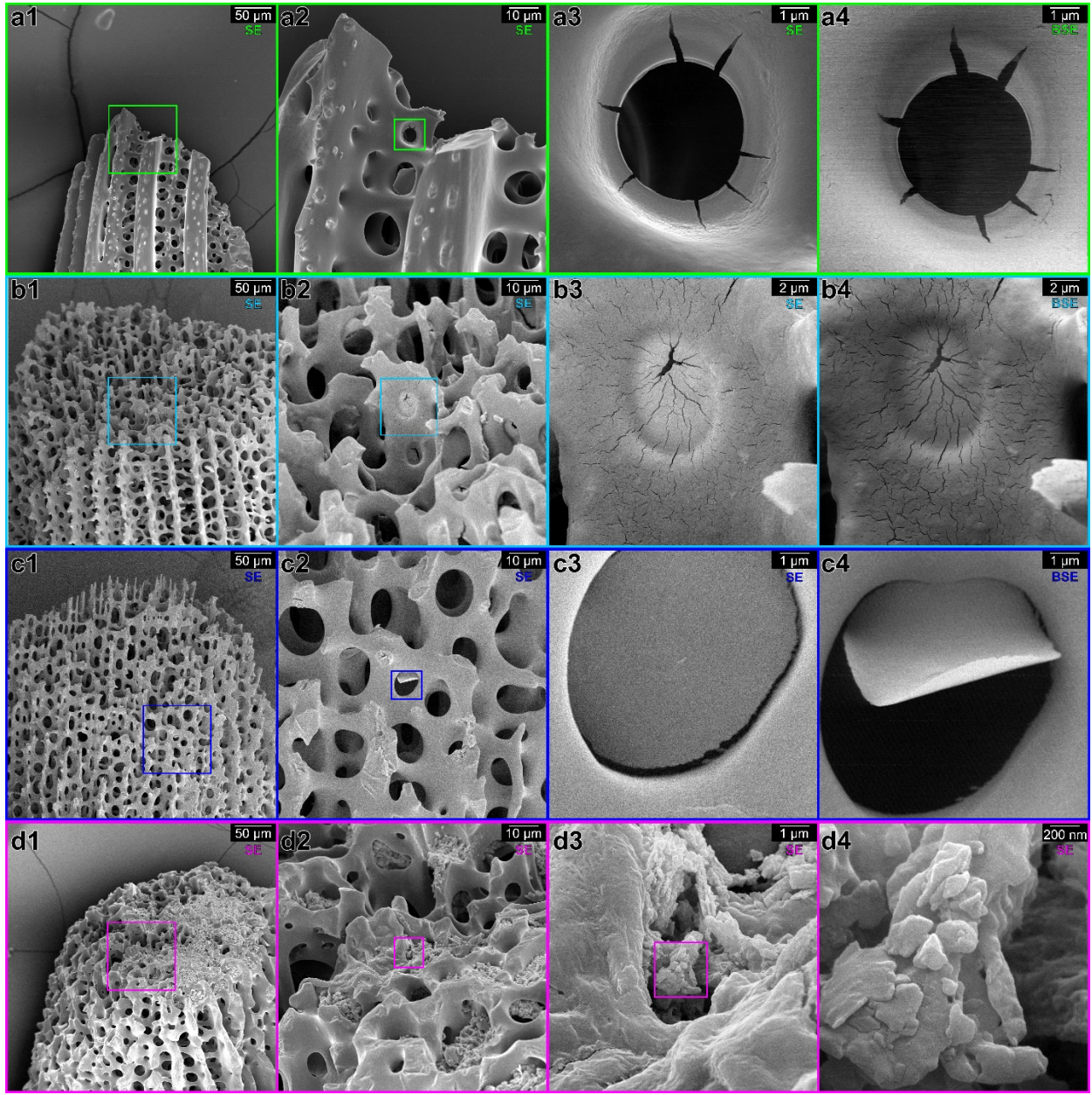

**Figure 4. More clues about the liquid-like precursor from 4 spines at increasing magnifications, imaged in SE or BSE**. a. This liquid-like pool had a circular opening when the precursor was liquid, then it cracked when the material either solidified, or crystallized, or got dehydrated in vacuum. b. This liquid-like pool and its surrounding stereom are both covered and cracked. One interpretation is that the liquid precursor coated both, then it shrank in volume

when it solidified, or crystallized, or was dehydrated. c. This liquid-like pool shows damage with exposure to the e-beam and was seen moving during the experiment. c3 was acquired in SE before electron-induced motion, c4 was acquired in BSE after the motion. Notice how thin the liquid-like pool is. d. This regenerating spine's stereom holes are not filled with liquid but with solid particles. These particles, therefore, must have solidified before falling into the stereom holes. They were, presumably, within the tissue when it was beached.

**Supporting Information**.

SI.pdf containing Figures S1-S7 and Materials and Methods (PDF)

C1_Cni5_ACCH2O.txt, C2_Cni5_ACC.txt, C3_Cni5_Calcite.txt (text files)

AUTHOR INFORMATION

**Corresponding Author**

* Pupa Gilbert: pupa@physics.wisc.edu

**Author Contributions**

PG conceived the initial experiments. CEK and CAS prepared all the samples, PG collected all the SEM data, CAS and PG collected PEEM data. PG organized all the data into figures and wrote the manuscript first draft, all co-authors edited the manuscript. Authors declare no competing interests.

**Funding Sources**

PG acknowledges 80% support from the U.S. Department of Energy, Office of Science, Office of Basic Energy Sciences, Chemical Sciences, Geosciences, and Biosciences Division, under Award DE-FG02-07ER15899, 10% support from NSF grant DMR-1603192, and 10% support from the Laboratory Directed Research and Development (LDRD) program at Berkeley Lab, through DOE-BES, under Award Number DE-AC02-05CH11231. PEEM experiments were done at the Advanced Light Source, which is supported by the Director, Office of Science, Office of Basic Energy Sciences, US Department of Energy under Contract No. DE-AC02-05CH11231.

ACKNOWLEDGMENTS

We thank Karl Menard and his colleagues at Bodega Marine Laboratory for collecting and providing the living sea urchins, Erol Kepkep and his colleagues for housing the sea urchins in their tanks at U.C. Berkeley during the spine regeneration experiments, Chang-Yu Sun for fixing and polishing the sea urchin spine for PEEM, and Hans Bechtel for FTIR experiments (not included), and Benjamin I. Fordyce for writing MATLAB code. CEK works in the lab of Drs. John Gerhart and Fred Wilt at U.C. Berkeley. We thank Yael Politi, Oliver Spaeker, Luca Bertinetti, and Helmut Cölfen for discussions.

REFERENCES

1. Knoll, A. H., Biomineralization and evolutionary history. *Revs Mineral Geochem* **2003,** *54* (1), 329-356.
2. Lowenstam, H. A., Minerals formed by organisms. *Science* **1981,** *211* (4487), 1126-1131.
3. Munch, E.; Launey, M. E.; Alsem, D. H.; Saiz, E.; Tomsia, A. P.; Ritchie, R. O., Tough, bio-inspired hybrid materials. *Science* **2008,** *322* (5907), 1516-1520.
4. Lowenstam, H. A.; Weiner, S., *On biomineralization*. Oxford University Press on Demand: 1989.
5. Beniash, E.; Aizenberg, J.; Addadi, L.; Weiner, S., Amorphous calcium carbonate transforms into calcite during sea urchin larval spicule growth. *Procs Natl Acad Sci* **1997,** *264* (1380), 461-465.
6. Politi, Y.; Arad, T.; Klein, E.; Weiner, S.; Addadi, L., Sea urchin spine calcite forms via a transient amorphous calcium carbonate phase. *Science* **2004,** *306* (5699), 1161-1164.
7. De Yoreo, J. J.; Gilbert, P. U. P. A.; Sommerdijk, N. A. J. M.; Penn, R. L.; Whitelam, S.; Joester, D.; Zhang, H.; Rimer, J. D.; Navrotsky, A.; Banfield, J. F.; Wallace, A. F.; Michel, F. M.; Meldrum, F. C.; Cölfen, H.; Dove, P. M., Crystallization by particle attachment in synthetic, biogenic, and geologic environments. *Science* **2015,** *349* (6247), aaa6760.
8. Gower, L. B.; Odom, D. J., Deposition of calcium carbonate films by a polymer-induced liquid-precursor (PILP) process. *J Cryst Growth* **2000,** *210* (4), 719-734.

9. Demichelis, R.; Raiteri, P.; Gale, J. D.; Quigley, D.; Gebauer, D., Stable prenucleation mineral clusters are liquid-like ionic polymers. *Nat Comms* **2011**, *2*, 590.
10. Gebauer, D.; Völkel, A.; Cölfen, H., Stable prenucleation calcium carbonate clusters. *Science* **2008**, *322* (5909), 1819-1822.
11. Gower, L. B., Biomimetic model systems for investigating the amorphous precursor pathway and its role in biomineralization. *Chem Revs* **2008**, *108* (11), 4551-4627.
12. Avaro, J. T.; Wolf, S. L.; Hauser, K.; Gebauer, D., Stable Prenucleation Calcium Carbonate Clusters Define Liquid–Liquid Phase Separation. *Angew Chem* **2020**, *59* (15), 6155-6159.
13. Rodriguez-Navarro, C.; Ruiz-Agudo, E.; Harris, J.; Wolf, S. E., Nonclassical crystallization in vivo et in vitro (II): Nanogranular features in biomimetic minerals disclose a general colloid-mediated crystal growth mechanism. *J Struct Biol* **2016**, *196* (2), 260-287.
14. Schenk, A. S.; Zope, H.; Kim, Y.-Y.; Kros, A.; Sommerdijk, N. A.; Meldrum, F. C., Polymer-induced liquid precursor (PILP) phases of calcium carbonate formed in the presence of synthetic acidic polypeptides—relevance to biomineralization. *Faraday Discuss* **2012**, *159* (1), 327-344.
15. Bather, F., Stereom. *Nature* **1891**, *43* (1111), 345-345.
16. Aizenberg, J.; Tkachenko, A.; Weiner, S.; Addadi, L.; Hendler, G., Calcitic microlenses as part of the photoreceptor system in brittlestars. *Nature* **2001**, *412* (6849), 819-822.
17. Donnay, G.; Pawson, D. L., X-ray diffraction studies of echinoderm plates. *Science* **1969**, *166* (3909), 1147-1150.
18. Finnemore, A. S.; Scherer, M. R.; Langford, R.; Mahajan, S.; Ludwigs, S.; Meldrum, F. C.; Steiner, U., Nanostructured calcite single crystals with gyroid morphologies. *Adv Mater* **2009**, *21* (38-39), 3928-3932.
19. Albéric, M.; Stifler, C. A.; Zou, Z.; Sun, C.-Y.; Killian, C. E.; Valencia Molina, S.; Mawass, M.-A.; Bertinetti, L.; Gilbert, P. U. P. A.; Politi, Y., Growth and regrowth of adult sea urchin spines involve hydrated and anhydrous amorphous calcium carbonate precursors. *J Struct Biol X* **2019**, *1*, 1000004.
20. Weiss, I. M.; Tuross, N.; Addadi, L.; Weiner, S., Mollusc larval shell formation: amorphous calcium carbonate is a precursor phase for aragonite. *J Exper Zool* **2002**, *293* (5), 478-491.
21. Vidavsky, N.; Masic, A.; Schertel, A.; Weiner, S.; Addadi, L., Mineral-bearing vesicle transport in sea urchin embryos. *Journal of structural biology* **2015**, *192* (3), 358-365.
22. Kahil, K.; Varsano, N.; Sorrentino, A.; Pereiro, E.; Rez, P.; Weiner, S.; Addadi, L., Cellular pathways of calcium transport and concentration toward mineral formation in sea urchin larvae. *Procs Natl Acad Sci* **2020**, *117* (49), 30957-30965.
23. Cheng, X.; Gower, L. B., Molding Mineral within Microporous Hydrogels by a Polymer-Induced Liquid-Precursor (PILP) Process. *Biotechnology progress* **2006**, *22* (1), 141-149.
24. Kim, Y.-y.; Gower, L. B., Formation of complex non-equilibrium morphologies of calcite via biomimetic processing. *MRS Online Proceedings Library* **2003**, *774* (1), 671-678.
25. Gong, Y. U. T.; Killian, C. E.; Olson, I. C.; Appathurai, N. P.; Amasino, A. L.; Martin, M. C.; Holt, L. J.; Wilt, F. H.; Gilbert, P. U. P. A., Phase transitions in biogenic amorphous calcium carbonate. *Procs Natl Acad Sci* **2012**, *109*, 6088-6093.
26. Gilbert, P. U.; Porter, S. M.; Sun, C.-Y.; Xiao, S.; Gibson, B. M.; Shenkar, N.; Knoll, A. H., Biomineralization by particle attachment in early animals. *Procs Natl Acad Sci* **2019**, *116*, 17659–17665.

27.	Sun, C.-Y.; Stifler, C. A.; Chopdekar, R. V.; Schmidt, C. A.; Parida, G.; Schoeppler, V.; Fordyce, B. I.; Brau, J. H.; Mass, T.; Tambutté, S.; Gilbert, P. U. P. A., From particle attachment to space-filling coral skeletons *Procs Natl Acad Sci* **2020,** *in press*.
28.	Morgulis, M.; Gildor, T.; Roopin, M.; Sher, N.; Malik, A.; Lalzar, M.; Dines, M.; de-Leon, S. B.-T.; Khalaily, L.; de-Leon, S. B.-T., Possible cooption of a VEGF-driven tubulogenesis program for biomineralization in echinoderms. *Proceedings of the National Academy of Sciences* **2019,** *116* (25), 12353-12362.
29.	Killian, C. E.; Wilt, F. H., Endocytosis in primary mesenchyme cells during sea urchin larval skeletogenesis. *Exper Cell Res* **2017,** *359* (1), 205-214.
30.	Kahil, K.; Varsano, N.; Sorrentino, A.; Pereiro, E.; Rez, P.; Weiner, S.; Addadi, L., Cellular pathways of calcium transport and concentration toward mineral formation in sea urchin larvae. *Proceedings of the National Academy of Sciences* **2020,** *117* (49), 30957-30965.
31.	Frazer, B. H.; Gilbert, B.; Sonderegger, B. R.; De Stasio, G., The probing depth of total electron yield in the sub keV range: TEY-XAS and X-PEEM. *Surf Sci* **2003,** *537*, 161-167.
32.	Shin, Y.; Brangwynne, C. P., Liquid phase condensation in cell physiology and disease. *Science* **2017,** *357* (6357).
33.	Olszta, M.; Douglas, E.; Gower, L., Scanning electron microscopic analysis of the mineralization of type I collagen via a polymer-induced liquid-precursor (PILP) process. *Calc Tiss Int* **2003,** *72* (5), 583-591.
34.	Wolf, S. E.; Gower, L. B., In *New perspectives on mineral nucleation and growth: From solution precursors to solid materials*, Van Driessche, A. E.; Kellermeier, M.; Benning, L. G.; Gebauer, D., Eds. Springer: 2017.
35.	Kellermeier, M.; Gebauer, D.; Melero-García, E.; Drechsler, M.; Talmon, Y.; Kienle, L.; Cölfen, H.; García-Ruiz, J. M.; Kunz, W., Colloidal stabilization of calcium carbonate prenucleation clusters with silica. *Adv Funct Mater* **2012,** *22* (20), 4301-4311.
36.	Gebauer, D.; Cölfen, H., Prenucleation clusters and non-classical nucleation. *Nano Today* **2011,** *6* (6), 564-584.
37.	Gebauer, D.; Kellermeier, M.; Gale, J. D.; Bergström, L.; Cölfen, H., Pre-nucleation clusters as solute precursors in crystallisation. *Chem Soc Revs* **2014,** *43* (7), 2348-2371.
38.	Wallace, A. F.; Hedges, L. O.; Fernandez-Martinez, A.; Raiteri, P.; Gale, J. D.; Waychunas, G. A.; Whitelam, S.; Banfield, J. F.; De Yoreo, J. J., Microscopic evidence for liquid-liquid separation in supersaturated CaCO3 solutions. *Science* **2013,** *341* (6148), 885-889.
39.	Zou, Z.; Habraken, W. J.; Bertinetti, L.; Politi, Y.; Gal, A.; Weiner, S.; Addadi, L.; Fratzl, P., On the phase diagram of calcium carbonate solutions. *Adv Mater Interf* **2017,** *4* (1), 1600076.
40.	Killian, C. E.; Metzler, R. A.; Gong, Y. T.; Olson, I. C.; Aizenberg, J.; Politi, Y.; Wilt, F. H.; Scholl, A.; Young, A.; Doran, A.; Kunz, M.; Tamura, N.; Coppersmith, S. N.; Gilbert, P. U. P. A., The mechanism of calcite co-orientation in the sea urchin tooth. *J Am Chem Soc* **2009,** *131*, 18404-18409.
41.	DeVol, R. T.; Sun, C.-Y.; Marcus, M. A.; Coppersmith, S. N.; Myneni, S. C. B.; Gilbert, P. U. P. A., Nanoscale Transforming Mineral Phases in Fresh Nacre. *J Am Chem Soc* **2015,** *137* (41), 13325-13333.
42.	Mass, T.; Giuffre, A. J.; Sun, C.-Y.; Stifler, C. A.; Frazier, M. J.; Neder, M.; Tamura, N.; Stan, C. V.; Marcus, M. A.; Gilbert, P. U. P. A., Amorphous calcium carbonate particles form coral skeletons. *Procs Natl Acad Sci* **2017,** *114* (37), E7670-E7678.
43.	Nudelman, F.; Shimoni, E.; Klein, E.; Rousseau, M.; Bourrat, X.; Lopez, E.; Addadi, L.; Weiner, S., Forming nacreous layer of the shells of the bivalves Atrina rigida and Pinctada margaritifera: an environmental-and cryo-scanning electron microscopy study. *J Struct Biol* **2008,** *162* (2), 290-300.
44.	Macías-Sánchez, E.; Checa, A. G.; Willinger, M. G. In *The transport system of nacre components through the surface membrane of gastropods*, Key Eng Mater, Trans Tech Publ: 2015; pp 103-112.

# Supplementary Information for
Evidence for a liquid precursor to biomineral formation..

Cayla A. Stifler[1], Christopher E. Killian[2], Pupa U. P. A. Gilbert[1,3,4†*].

1 Department of Physics, University of Wisconsin, Madison, WI 53706, USA.
2 Department of Molecular and Cell Biology, University of California, Berkeley, CA 94720, USA.
3 Departments of Chemistry, Geoscience, Materials Science, University of Wisconsin, Madison, WI 53706, USA.
4 Lawrence Berkeley National Laboratory, Chemical Sciences Division, Energy Geoscience Division, Berkeley, CA 94720, USA.
† previously publishing as Gelsomina De Stasio.
* Corresponding author: Pupa Gilbert

Email: pupa@physics.wisc.edu

**This PDF file includes:**

    Figures S1 to S7, Materials and Methods

**Other supplementary materials for this manuscript include the following:**

    "Cni5" component spectra

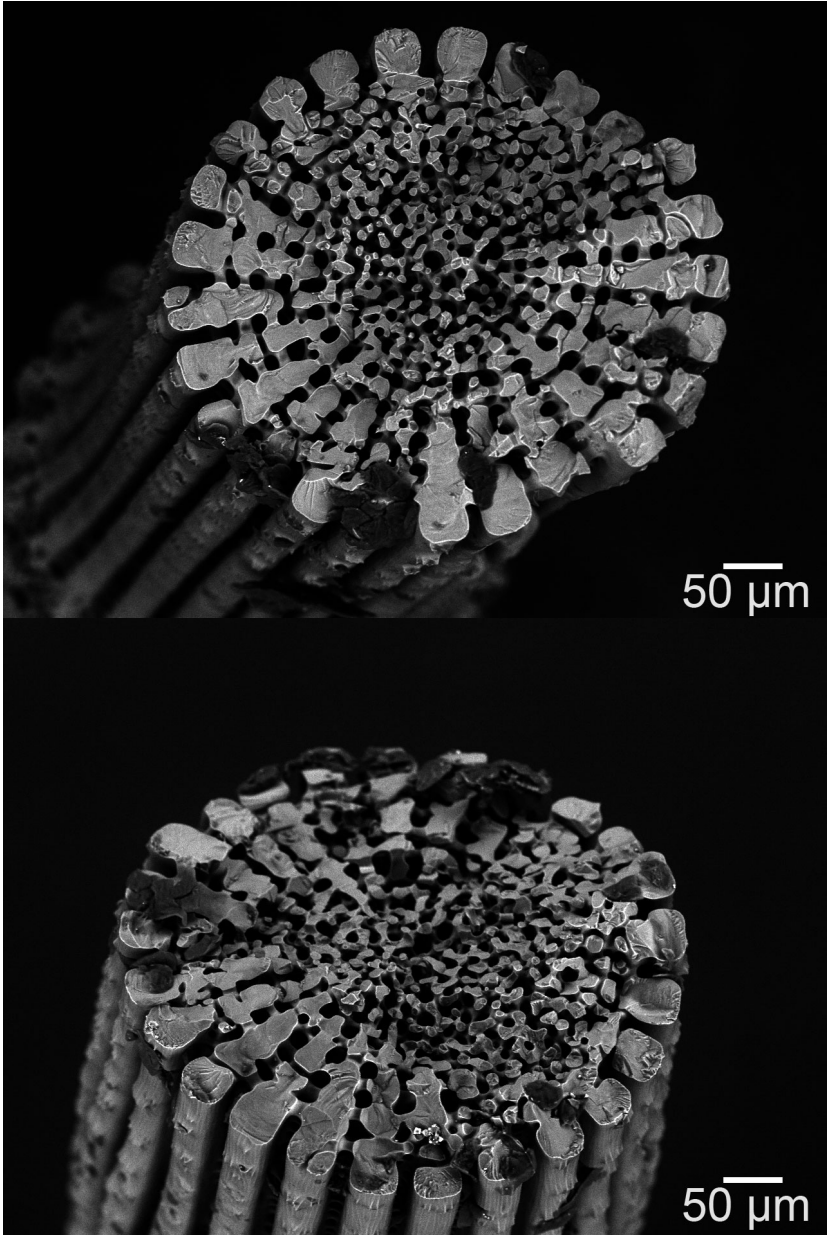

**Fig. S1. Bleached and fractured sea urchin spines**. The two spines from *Strongylocentrotus purpuratus* show that only holey stereom is present at the center. The radial structures visible at the outer diameters are termed septa and confer the striations visible on the lateral surface of the spine cylindrical structure.

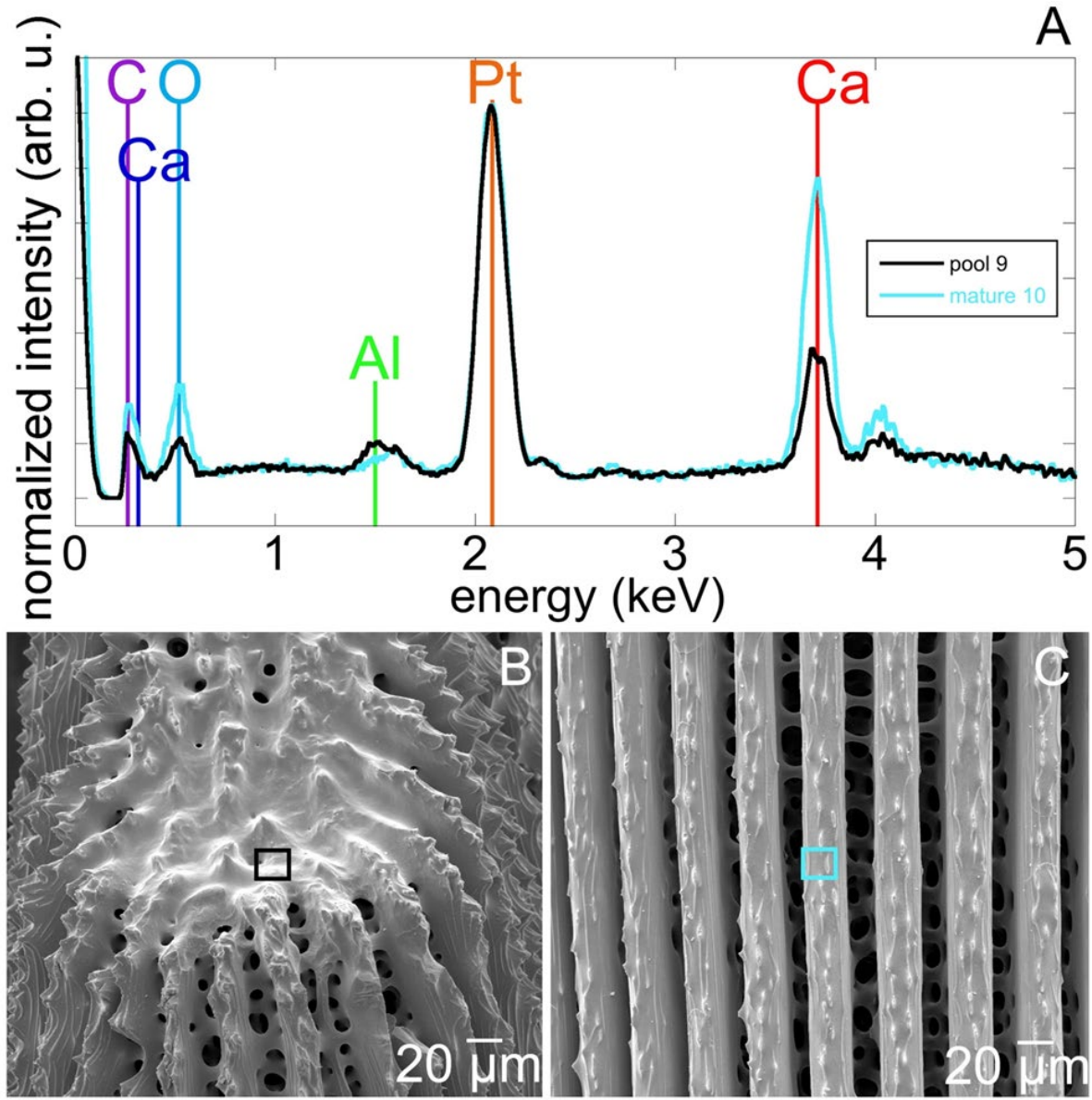

**Fig. S2.** Energy-dispersive x-ray (EDX) fluorescence emission data of the liquid-like pool and the mature spine. A. The two EDX spectra are similar but not identical. The Ca, O, C peak heights and therefore concentrations are greater in the mature-spine calcite (cyan) than in the liquid-like pool (black). Their relative decreases, however, differ. The approximate ratios after background subtraction, are as follows: [Ca pool]: [Ca mature]=0.4, [O pool]: [O mature]=0.4, and [C pool]: [C mature]=0.7. The Pt coating is identical in the two regions, despite topographic differences, because the 20 nanometers of Pt were deposited while tilting and spinning the sample. The ratio confirms that the Pt concentrations are identical, after background subtraction: [Pt pool]: [Pt mature]=1. B, C. SEM image acquired at 500x magnification, indicating the pool or mature regions from which the EDX spectra in A were acquired (black box in B, cyan box in C). Both spectra were acquired in the boxed regions at 10,000x magnification.

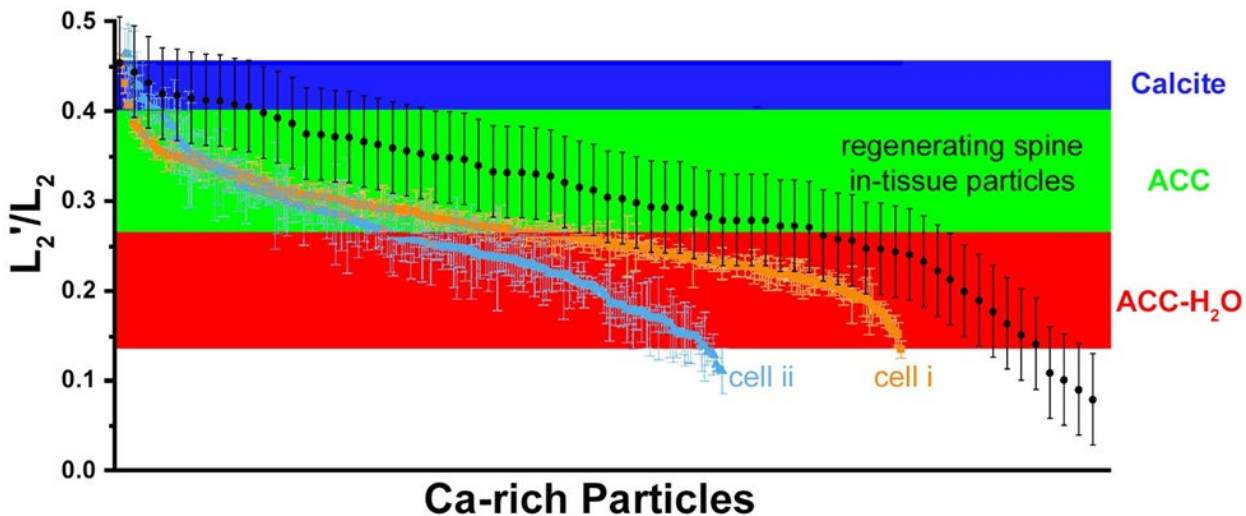

**Fig. S3.** Short-range order of Ca-rich particles in sea urchin embryo single-cells from Kahil et. al.[1] (cells i and ii) and in-tissue particles from regenerating adult sea urchin spine (this work). The short-range order is defined as the first or second nearest neighboring atoms to the Ca atoms in the observed phase, measured by the peak intensity ratio $L_2'/L_2$ as described by Kahil et. al.[1]. In their work or ours, pixel sizes were 13 nm or 55 nm, Ca L-edge spectra were acquired with 18 or 121 energy data points. We both minimized radiation damage by limiting the exposure time but made slightly different choices: Kahil et. al.[1] reduced the number of energy steps, we limited the spatial resolution and therefore the exposure time. The two approaches are completely equivalent because particle sizes are well above pixel sizes, thus it does not matter if they were observed in transmission or photoemission (their work and ours, respectively). The blue, green, and red bands in the background represent the peak ratio ranges for the three phases: calcite, ACC, and ACC-$H_2O$, respectively. The upper or lower bound of each band is the peak ratio of their work or our work, respectively. These differences in peak ratios are due to differences in beamline resolution and energy sampling. Peak ratios from the Ca L-edge spectra collected by PEEM in each pixel of in-tissue particles in a regenerating sea urchin spine (black data points, this work). Our in-tissue particle data points are superimposed on their single-cell data points. Only a subset of our data is shown, selected among the lowest noise spectra, which yield the most reliable peak ratios. In almost all in-tissue particles the $L_2'/L_2$ ratios are within the calcite, ACC, or ACC-$H_2O$ bands, indicating that at the time of analysis in PEEM these particles were solid. A few in-tissue particle pixels, however, were below the ACC-$H_2O$ band, suggesting greater disorder.

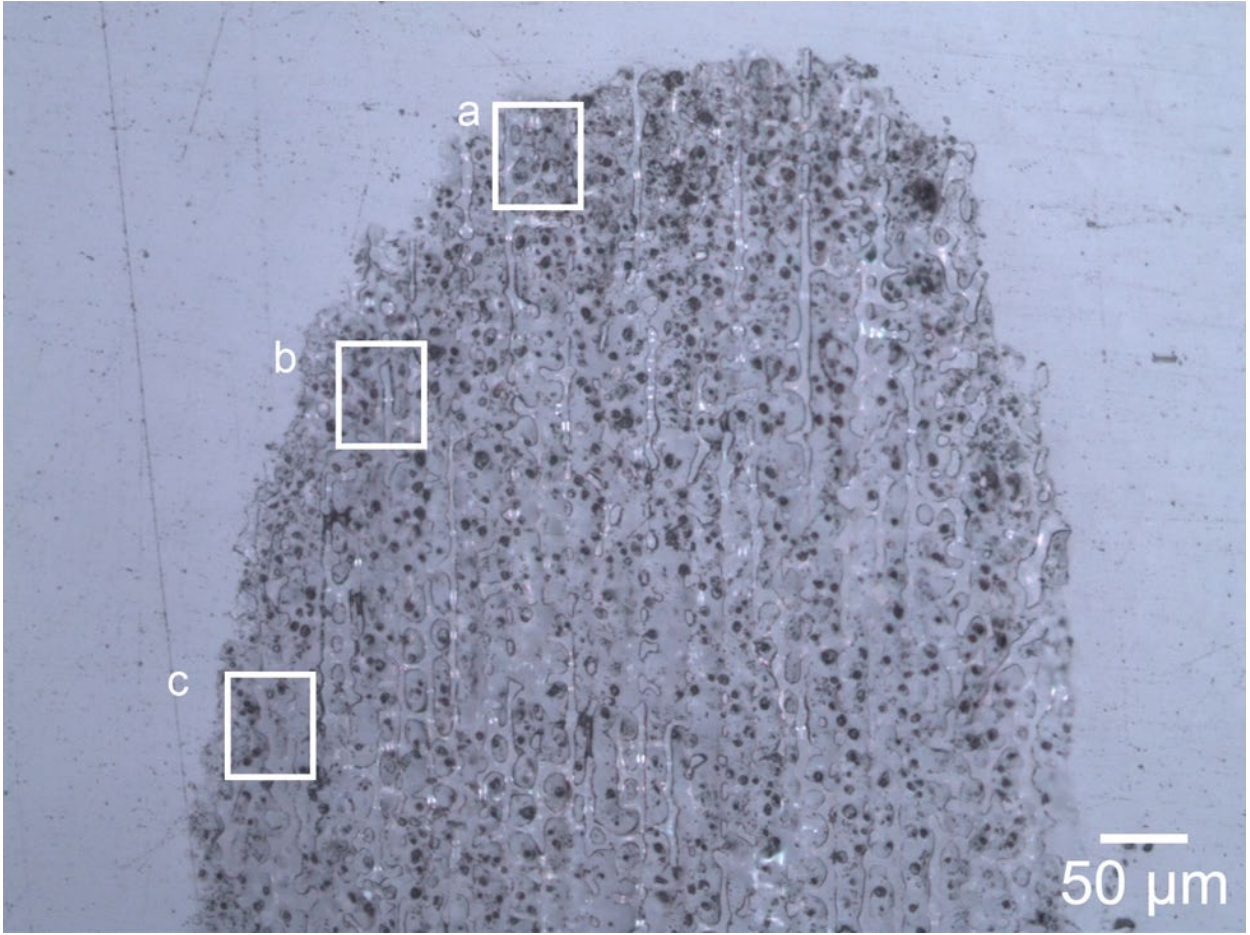

**Fig. S4. Visible light micrograph of fixed, regenerating spine.** The three boxed regions a, b, and c correspond to the identically labeled areas in Figure 3. The dark circles are bubbles in the embedding epoxy.

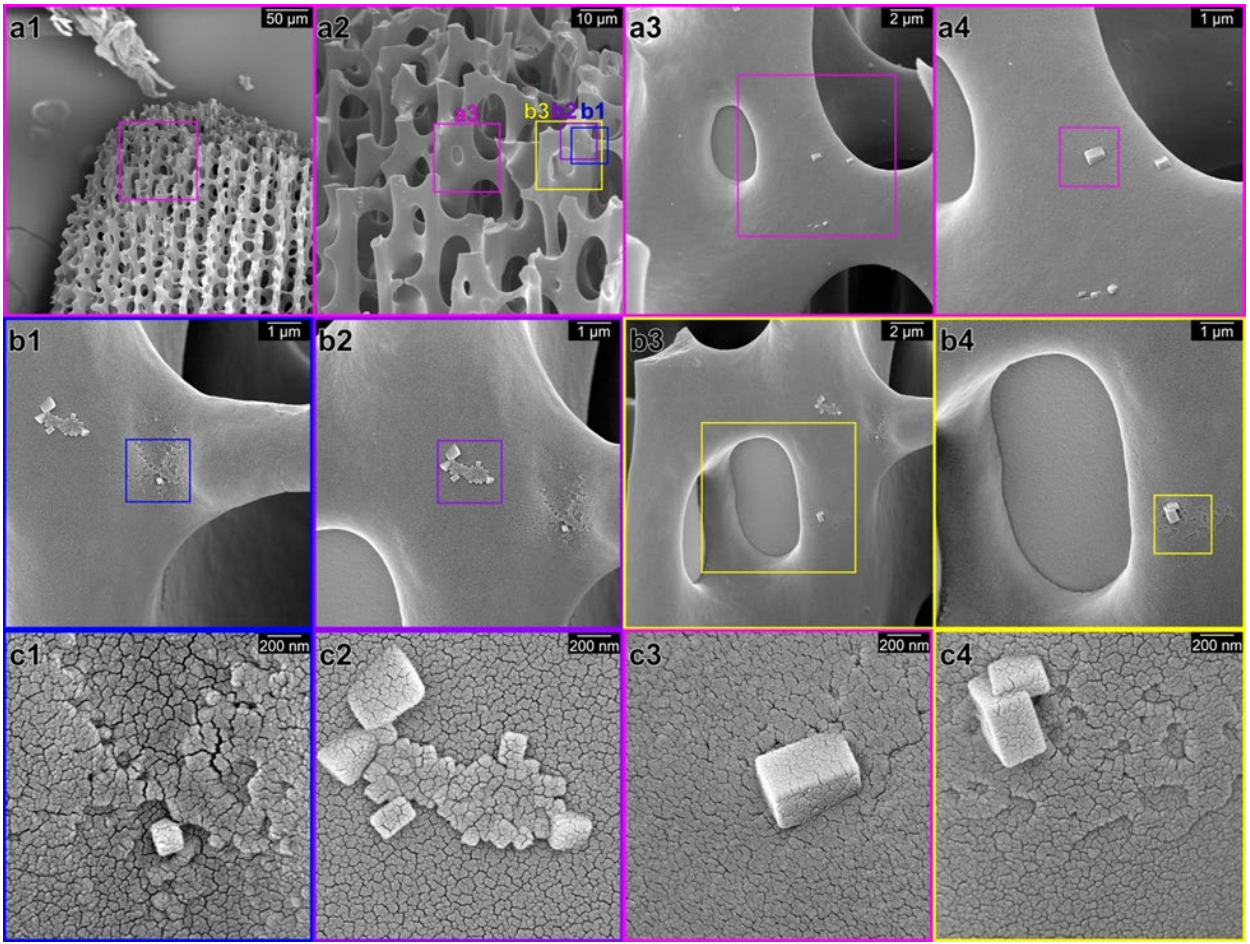

**Fig. S5. Rhombohedra on a sea urchin spine stereom**. All images were taken in SEM-SE mode. The rhombohedra show that the liquid-like precursor may crystallize as calcite. In Fig. S6 rhombohedra in BSE are as bright as or brighter than the stereom, whereas salts are darker. The rhombohedra, therefore, are likely calcite.

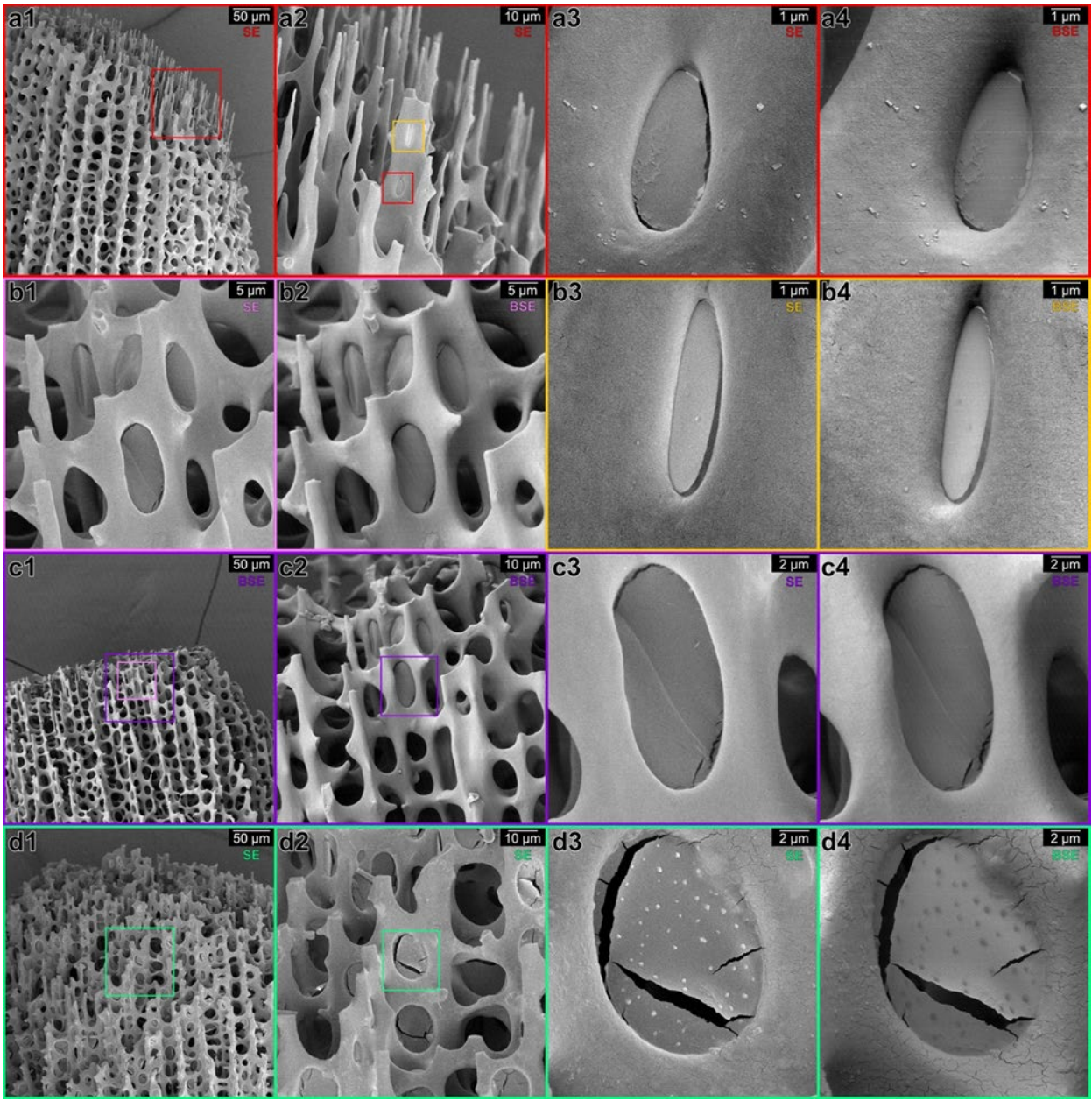

**Fig. S6.** Liquid-like pools in three regenerating spines from different animals, imaged in secondary electron (SE) or back-scattered electron (BSE) modes of SEM. SE and BSE labels appear in each panel. Boxes in a1,a2 or c1,c2 indicate regions magnified in a2,a3,a4,b3,b4, or in c2,c3,b1,b2 and correspondingly colored. b1,b2 show three liquid-like pools in the same frame. c1 shows more liquid-like pools at lower magnification. Notice that the pools in a4,b4,c4 are as bright as the rest of the stereom in BSE, indicating that the Ca density is similar. The pool in a3,a4 is surrounded by submicron rhombohedra bright in both SE and BSE. The pool in d3,d4 is covered and surrounded by salt crystals, which are bright in SE but dark in BSE. The rhombohedra in a3,a4, therefore, must be calcite. This indicates that the liquid-like material crystallizes as calcite.

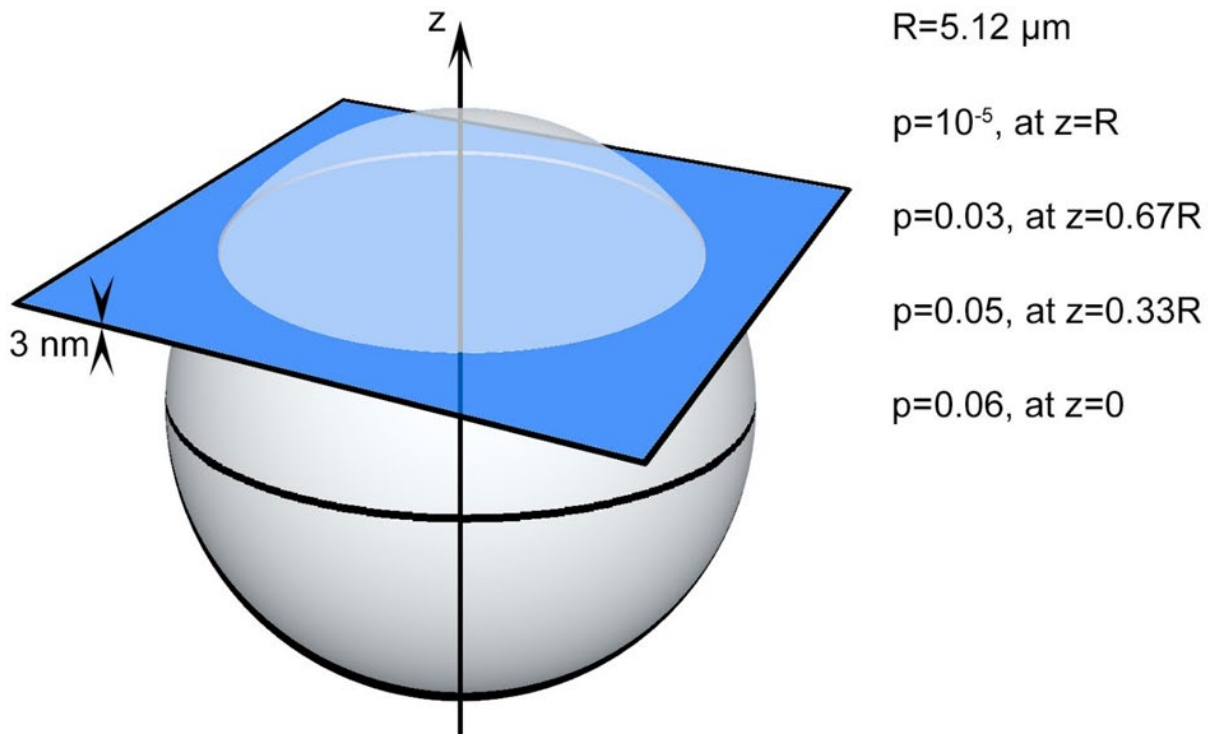

**Fig. S7.** A spherical model of a calcifying cell showing, on the right, the probability p that a 3nm-thick slice (the probing depth of PEEM at the Ca L-edge 31) of the cell at different heights on the sphere's vertical axis contains at least one Ca-rich vesicle. We assumed that there are 140 particles, each with 250nm-diameter, within each cell, as observed by Kahil et al.[1]. The probability of finding a Ca-rich particle in the 3nm-thick slice analyzed in PEEM is even lower, because the 55μm-field of view of PEEM images may or may not contain a 10.24μm-diameter stereom-forming cell, or a portion of a cell.

**Materials and Methods**

**Sea urchins, spine regeneration**

Twenty-five *Strongylocentrotus purpuratus* sea urchins were collected near Bodega Bay, CA, USA. They were then kept in outdoors tanks with seawater circulating, coming directly from the Pacific Ocean at the University of California - Bodega Marine Laboratory, Bodega, CA, USA. They were then transferred to UC-Berkeley, indoors, kept at 15 °C constant temperature in seawater, and fed kelp every day for ~ 1 week before the experiments to ensure that the urchins were healthy: they moved around and had their tube feet out. Several spines in each sea urchin were cut off at approximately ½ of their length using scissors. The spines were allowed to regenerate for 7-14 days and harvested. Individual spines were selected if they appeared to be regenerating. These regenerating spines were harvested by cutting them near the base.

A total of 28 spines were prepared vertically and 2 horizontally for SEM experiments and 1 vertically for PEEM experiments. Horizontally prepared spines showed liquid-like pools in 0 out of 2 spines; thus, the horizontal-spine preparation was abandoned. Vertically prepared spines showed liquid-like pools in 20 out of 28 spines. Therefore, only this procedure was used subsequently.

**Sample preparations for SEM experiments**

Each spine was placed vertically into a 1.5 ml Eppendorf tube, with the regenerating tip up.

The spines were treated, on ice, as follows:

- Two rinses with 50 mM Tris base at pH 9.
- Two incubations for 10 min each in 6% NaClO.
- Two rinses with 50 mM Tris base at pH 9.
- Two rinses with 100% ethanol.
- Air dried at room temperature.

For SEM experiments the spines were mounted on carbon tape (Ted Pella Inc., Redding, CA), and coated with 20-nm Pt, while rotating and tilting the samples, using a high-resolution Cressington 208hr sputter coater with rotary, planetary, and tilting movements (Cressington, UK), which is optimized for coating distribution and coverage of highly topographic samples such as sea urchin spines.

**Sample preparations for PEEM experiments**
One of the regenerating spines, harvested 10 days after cutting, was placed vertically in a tube of seawater with 5% $MgCl_2$ and subsequently fixed with formaldehyde in 0.05 M sodium cacodylate buffer and gradually dehydrated in five steps with ethanol concentrations increasing from 50% to 90% with each 10-minute step. Finally, the spine was dehydrated in 100% ethanol twice, for 10 minutes each time. The spine was then embedded in Solarez resin and cured according to the procedure in Sun et. al. 2020[2]. Once cured, the sample was polished and coated with 1 nm Pt in the area of interest and 40 nm elsewhere while spinning and tilting.

**SEM experiments**

Most of the SEM experiments were all done at the UC-Berkeley Electron Microscopy Laboratory, using either a Hitachi TM1000 or a Hitachi S5000 microscope.

The TM1000 has a fixed accelerating voltage of 15 kV, fixed back-scattered electron (BSE) mode, and 10,000x maximum magnification. This was used for Fig. 1b-f, Fig. 2b4, and Fig. S1. The S5000 was used with 10 kV accelerating voltage, in both secondary electrons (SE) and BSE modes, and has a 1,000,000x maximum magnification. For the present experiments the maximum magnification used was 50,000x. Each region of interest was first images at 50,000x, then at 10,000x, 5,000x, 1,000x, and 250x. The S5000 was use for Figs. 2, 4, S5, S6.

In all figures, the images were presented in reverse order, with low magnification first. All SEM images were adjusted for brightness and contrast in Adobe Photoshop® and assembled with colored grids and boxes also in Photoshop®.

Energy-Dispersive X-ray (EDX) fluorescence emission analysis was done at Lawrence Berkeley National Laboratory, Chemical Sciences Division, building 30, using an FEI Quanta FEG 250 SEM, equipped with a Bruker EDX analyzer, and used with a 5 kV e-beam and 10 mm working distance.

**PEEM experiments**
PEEM experiments were done at the Advanced Light Source on beamline 11.0.1.1 at the Ca L-edge for component mapping with a lateral resolution of 60 nm and the probing depth 3 nm. All data from 3 areas of the same regenerating spine were acquired 30, 30.5, and 31 hours after extraction, to ensure that the amorphous phases stayed amorphous, as much as possible, considering the unavoidably long sample preparation time (30 hours). The spine was not bleached, but fixed as described in [2] to retain spine tissues and in-tissue particles. All data were acquired on the regenerating part of the spine. We limited the magnification and resolution (55 nm instead of the possible 20 nm or even 10 nm), to shorten the acquisition time and thus reduce the radiation dose on the spine to minimize radiation damage and maintain the phases amorphous. For component maps, stacks of PEEM images were acquired while scanning the illuminating x-ray photon energy across the Ca L-edge, from 340 to 360 eV, with 0.1-eV energy steps between 345 and 355 eV and 0.5 eV from 340 to 345 eV and 355 to 360 eV (121 energy steps). All Ca data were taken with circular polarization to avoid any linear polarization effects from crystals. The 121 images were converted to 8-bit tif stacks using Igor Pro® version 6.37, and the GG Macros developed for component analysis by our group, and available to any interested users free of charge (https://home.physics.wisc.edu/gilbert/software/). The component spectra used were obtained by averaging single-pixel spectra containing at least 90% of the relevant phase, in sea urchin spine data from Alberic et al. 2019[3], but improved the component spectra compared with that publication. The new component spectra, called "Cni5", improved the peak-fitting of the spectra, with identical backgrounds for all 3 phases, thus background variations potentially leading to misinterpretation were eliminated. The Cni5 spectra are included as text files. The results were reproduced on 3 diverse areas (Fig. 3) of the same regenerating spine. In component maps, the skeleton and the calcium containing particles in the tissue were preserved, and the rest of the pixels were masked (black) with a series of overlapping masks. One mask removed all pixels with a bad fit to the 3 component spectra, defined as one with too high a chi-square value: $c^2 > 0.01$, or 0.015, or 0.025 for areas 1, 2, or 3, respectively (Igor and GG Macros). A second mask called "Therm mask" removed pixels where the phase went thermodynamically uphill in the repeat movie (e.g. a green pixel becoming red, which is a non-physical result) with an allowed tolerance of 100 color levels out of 255 for areas 1 and 3 and allowed tolerance of 25 color levels out of 255 for the much noisier area 2 (MATLAB®, MathWorks). A third mask called "magic eraser" removed particles smaller than 2 pixels by 2 pixels, interpreted as noise that escaped the first 2 masks. The magic eraser places each pixel in the center of a 3x3 grid comprised of 8 surrounding pixels and masks it if the pixel does not border at least 3 other unmasked pixels (MATLAB®, MathWorks). All remaining pixels clustered into reliable particles in the tissue. A fourth mask was done by hand: single-pixel spectra from in-tissue particles were carefully examined, and any pixel whose spectrum was mostly noise was masked off by hand. The peak intensity ratios in Fig. S3 were obtained by first subtracting a linear fit to the pre-edge of each spectrum, and then measuring the intensity of peaks $L_2$' and $L_2$, previously termed peaks 2 and 1, respectively [3-6]. The background subtraction and peak intensity measurements were done in Igor Pro 6.37, using the GG Macros. The ratios were calculated in Microsoft Excel® for Mac, plotted in Kaleidagraph® 4.5.2, and superimposed on the Kahil et al. figure using Adobe Photoshop® version 22.2.0 for Mac.

**Probability calculations**
The calcifying cell was modeled as a sphere (Fig. S7) with a radius of 5.12 µm based on the dimensions of the cell shown in Kahil et. al. [1]. The Ca-rich particles were modeled as smaller spheres with a diameter of 250 nm. To find the total number of Ca-rich vesicles that could fit inside the cell, the volume of the cell was divided by the volume of the particle and that ratio was multiplied by 0.63, the packing efficiency of randomly packed spheres[7]. Then, to find the probability that a single "spot" is filled with a vesicle, the average number of vesicles per cells from Kahil et. al. [1], 140, was divided by the total number of spots, 43,293. The sphere was then divided into 3nm-thick sections to represent the probing depth of PEEM [8]. The number of "spots" in a given section was calculated from the division of the volume of the spherical segment [9] by the volume of a 250nm vesicle. The probability that no particles are found at a given spot is then one minus the probability of finding a particle in that spot. From there, the probability that the spots in the 3nm slice contains no vesicles was calculated by the probability of a single spot having no particles raised to the number of vesicles in that slice. The probability that there is at least one particle in a given slice is one minus the probability of having no particles in that slice.

**Cni5 component spectra (separate files).** C1_Cni5_ACCH2O.txt, C2_Cni5_ACC.txt, and C3_Cni5_Calcite.txt

**SI References**

1. Kahil, K.; Varsano, N.; Sorrentino, A.; Pereiro, E.; Rez, P.; Weiner, S.; Addadi, L., Cellular pathways of calcium transport and concentration toward mineral formation in sea urchin larvae. *Procs Natl Acad Sci* **2020,** *117* (49), 30957-30965.
2. Sun, C.-Y.; Stifler, C. A.; Chopdekar, R. V.; Schmidt, C. A.; Parida, G.; Schoeppler, V.; Fordyce, B. I.; Brau, J. H.; Mass, T.; Tambutté, S.; Gilbert, P. U. P. A., From particle attachment to space-filling coral skeletons *Procs Natl Acad Sci* **2020,** *in press*.
3. Albéric, M.; Stifler, C. A.; Zou, Z.; Sun, C.-Y.; Killian, C. E.; Valencia Molina, S.; Mawass, M.-A.; Bertinetti, L.; Gilbert, P. U. P. A.; Politi, Y., Growth and regrowth of adult sea urchin spines involve hydrated and anhydrous amorphous calcium carbonate precursors. *J Struct Biol X* **2019,** *1*, 1000004.
4. DeVol, R. T.; Sun, C.-Y.; Marcus, M. A.; Coppersmith, S. N.; Myneni, S. C. B.; Gilbert, P. U. P. A., Nanoscale Transforming Mineral Phases in Fresh Nacre. *J Am Chem Soc* **2015,** *137* (41), 13325-13333.
5. Gong, Y. U. T.; Killian, C. E.; Olson, I. C.; Appathurai, N. P.; Amasino, A. L.; Martin, M. C.; Holt, L. J.; Wilt, F. H.; Gilbert, P. U. P. A., Phase transitions in biogenic amorphous calcium carbonate. *Procs Natl Acad Sci* **2012,** *109*, 6088-6093.
6. Politi, Y.; Metzler, R. A.; Abrecht, M.; Gilbert, B.; Wilt, F. H.; Sagi, I.; Addadi, L.; Weiner, S.; Gilbert, P. U. P. A., Transformation mechanism of amorphous calcium carbonate into calcite in the sea urchin larval spicule. *Procs Natl Acad Sci* **2008,** *105* (45), 17362-17366.
7. Song, C.; Wang, P.; Makse, H. A., A phase diagram for jammed matter. *Nature* **2008,** *453* (7195), 629-632.
8. Frazer, B. H.; Gilbert, B.; Sonderegger, B. R.; De Stasio, G., The probing depth of total electron yield in the sub keV range: TEY-XAS and X-PEEM. *Surf Sci* **2003,** *537*, 161-167.
9. Harris, J. W.; Stöcker, H., *Handbook of mathematics and computational science*. Springer Science & Business Media: 1998.